\documentclass[a4paper,fleqn]{cas-dc}

\usepackage[authoryear]{natbib}
\usepackage{url}
\usepackage{hyperref}
\usepackage{aas_macros}

\def\tsc#1{\csdef{#1}{\textsc{\lowercase{#1}}\xspace}}
\tsc{WGM}
\tsc{QE}
\tsc{EP}
\tsc{PMS}
\tsc{BEC}
\tsc{DE}
\newcommand{\farcs}{.\!\!^{\prime\prime}}

\begin{document}
\let\WriteBookmarks\relax
\def\floatpagepagefraction{1}
\def\textpagefraction{.001}
\shorttitle{Quenching and Bulge–Disk Growth at Cosmic Noon}
\shortauthors{Chiara Mancini et~al.}

\title [mode = title]{A SHARP Look at Quenching and Bulge–Disk Growth in Massive Galaxies at Cosmic Noon} 
                     
\tnotemark[1,2]

\author[1]{Chiara Mancini}[type=editor,
                        orcid=0000-0002-4297-0561]
\cormark[1]
\ead{chiara.mancini@inaf.it}

\credit{Writing - Original draft, Conceptualization, Methodology, Data curation}
\affiliation[1]{organization={INAF - Istituto di Astrofisica Spaziale e Fisica Cosmica, Milano},
                addressline={Via Alfonso Corti 12}, 
                city={Milano},
                postcode={20133}, 
                country={Italy}}

\author[1]{Adriana Gargiulo}\credit{Writing - review}
\author[2]{Alvio Renzini}\credit{Writing - review}
\author[3,4]{Emanuele Daddi}\credit{Writing - review}
\author[5]{Stefano Zibetti}\credit{Writing - review}
\author[6]{Fabio R. Ditrani}\credit{Writing - review}
\author[5]{Anna R. Gallazzi}\credit{Writing - review}
\author[1]{Susanna Bisogni}\credit{Writing - review}
\author[1]{Paolo Franzetti}\credit{Software}
\author[1]{Giustina Vietri}\credit{Writing - review}
\author[7]{Louis Gabarra}\credit{Writing - review}
\author[4]{Fabrizio Gentile}\credit{Writing - review}
\author[6]{Marcella Longhetti}\credit{Writing - review}
\author[1]{Marco Scodeggio}\credit{Writing - review}


\affiliation[2]{organization={INAF - Osservatorio Astronomico di Paodva},
                addressline={Vicolo dell'Osservatorio, 5}, 
                postcodesep={}, 
                city={Padova},
                postcode={35122}, 
                country={Italy}}


\affiliation[3]{organization={Universit\'e Paris-Saclay, Universit\'e Paris Cit\'e, CEA, CNRS, AIM},
                postcodesep={}, 
                city={Gif-sur-Yvette},
                postcode={35122}, 
                country={France}}

\affiliation[4]{organization={CEA Saclay, DFR/IRFU, Service d’Astrophysique}, 
                addressline={Bat. 709},
                postcodesep={}, 
                city={Gif-sur-Yvette},
                postcode={91191}, 
                country={France}}

\affiliation[5]{organization={INAF - Osservatorio Astronomico di Arcetri},
                addressline={Largo Enrico Fermi, 5}, 
                postcodesep={}, 
                city={Firenze},
                postcode={50126}, 
                country={Italy}}
                
\affiliation[6]{organization={INAF - Osservatorio Astronomico di Brera},
                addressline={Via Brera, 28}, 
                postcodesep={}, 
                city={Milano},
                postcode={20121}, 
                country={Italy}}

\affiliation[7]{organization={Department of Physics, Oxford University},
                addressline={Keble Road}, 
                postcodesep={}, 
                city={Oxford},
                postcode={OX1 3RH}, 
                country={UK}}

\cortext[cor1]{Corresponding author}

\begin{abstract} 
The physical mechanisms that quench star formation in massive galaxies remain poorly understood. At cosmic noon ($1\lesssim z\lesssim3$), when star formation and AGN activity peak, galaxies rapidly evolve from star-forming disks into quiescent, bulge-dominated systems. While quenching correlates with stellar mass and bulge growth, the causal link between bulge assembly, star-formation suppression, and feedback processes remains unclear. 
Stellar population analysis from spatially resolved spectroscopy of galaxies caught during quenching is crucial to advance on this issue.
We show that the SHARP--VESPER Integral-Field Spectrograph (IFS) can efficiently fill this gap by combining ELT-class sensitivity, 31~mas spatial resolution ($>$3$\times$ sharper than JWST) and broad near-IR wavelength coverage with 12-IFU multiplexing.
This will enable, for the first time, a simultaneous bulge–disk decomposition of stellar populations and spatially resolved mapping of ionised gas in massive galaxies (log~M$\ast$/M$\odot\geq11$) at $2.2<z<3.5$, targeting systems on the main sequence shortly before quenching or already in the green valley. With typical exposure times of $\sim$15~hr, we will obtain S/N$_{res}\geq$15–20, i.e., per spectral resolution element, on the (inner) bulge, and (outer) disk extracted spectral continuum, and S/N$_{res}$>5 for nebular lines ([O\,II], H$\beta$, [O\,III], H$\alpha$) on sub-kpc scales.
These observations will allow us to reconstruct independent bulge and disk star-formation histories, ages, metallicities, and $\alpha$-enhancements, while mapping spatially resolved star formation, gas kinematics, and feedback-driven outflows. By directly comparing the timing of bulge growth and star-formation suppression across galaxy components, this programme will test whether quenching proceeds inside-out, distinguish fast and slow quenching pathways, and link structural transformation to feedback processes in the most massive galaxies at cosmic noon. 
\end{abstract}

\maketitle
\section{Scientific Rationale}

\subsection{Galaxy Quenching and Structural Evolution }

\noindent In the current framework of galaxy evolution, galaxies build their stellar populations through the steady, continuous accretion of cold gas from their surrounding dark matter halos \citep{Binney_2004, Keres_2005, Dekel_2009a}. 
This process can sustain elevated star-formation rates (SFRs) over extended timescales, in contrast to more stochastic events such as major mergers. 
Consequently, it has been proposed as the mechanism underlying the observational evidence that, at all epochs, most galaxies form stars at a relatively steady rate, approximately proportional to their stellar mass, along the so-called Main Sequence (MS) of star-forming galaxies \citep[SFGs][and references therein]{Noeske_2007, Daddi_2007a, Elbaz_2007, Speagle_2014, Popesso_2023}.
Strong confirmation of this paradigm has come from the substantial investment of observing time with the Integral Field Spectrographs (IFS) at the VLT (e.g. SINFONI, KMOS). These observations revealed that most highly star-forming galaxies at cosmic noon ($z \sim 2$), the epoch when the Universe reached its peak star-formation activity \citep{Madau_2014}, are turbulent rotating disks, fragmented into kpc-scale star-forming clumps, yet showing no evidence of major mergers \citep{ForsterSchreiber_2006, ForsterSchreiber_2009, Wisnioski_2015, ForsterSchreiber_2018}. 

Although the broad picture of how galaxies assemble their mass across cosmic time is becoming better defined, several key aspects remain poorly understood.
One of the most elusive is the mechanism that halts star formation, transforming a galaxy from an active star-forming into a quiescent system. The puzzle is further complicated by the observational evidence that blue, star-forming galaxies, both in the local Universe and at high redshift, are predominantly disk-dominated, \citep[e.g., with S\'ersic indices near unity;][]{Sersic_1968}, whereas red, massive quiescent systems tend to be bulge-dominated, i.e. with markedly steeper light profiles, described by higher S\'ersic indices \citep{Trujillo_2001, Baldry_2006, vanderWel_2014, Wuyts_2014, Lin_2019, Huertas-Company2025arXiv}. These trends suggest that the cessation of star formation is linked to a morphological transformation, but the temporal sequence and relative timescales of quenching and structural evolution are also unclear. Galaxies in transition from an evolutionary stage to the other are also observed. Known as "green valley" galaxies \citep[][]{Salim_2014}, they show intermediate properties (e.g., colours, SFR, S\'ersic index) and lower number density compared to the quiescent and star forming populations.   

Statistically, the fraction of quiescent galaxies increases with both stellar mass and environmental density, leading to the definition of “mass quenching” and “environment quenching” as two distinct processes \citep{PengY_2010, Peng_2012}.
Environment quenching is due to external processes, more effective in dense regions, and on low-mass galaxies ($M_*<10^{10} M_{\odot}$), such as tidal interactions, and ram-pressure stripping caused by the hot intracluster medium \citep{Gunn_1972, Moore_1996, Vulcani_2020, Poggianti_2025}. In contrast, mass quenching is an internal process that should become efficient once galaxies reach a certain critical mass \citep{Birnboim_2003}. 
Several mechanisms have been proposed to explain the mass quenching, yet none has emerged as the clear  dominant \citep[see][]{Man_Belli_2018}. 

Among the most widely discussed possibilities is feedback from the central active galactic nucleus (AGN), which may either eject gas from the host galaxy \citep[][]{Granato_2004}  
or heat the circumgalactic medium, preventing the accretion of fresh cold gas needed to fuel star formation \citep[][]{Croton_2006}. 
This scenario is supported by the strong increase in AGN activity with galaxy mass, especially at $z \geq 1$ \citep{Reddy_2005, Fiore_2008, Brusa_2009, Rodighiero_2015}, as well as by observations of powerful AGN-driven outflows in massive galaxies \citep{Genzel_2014, ForsterSchreiber_2014, ForsterSchreiber_2019}. 

Another option is the morphological quenching, in which the growth of a massive central bulge stabilizes the disk and suppresses the formation of star-forming clumps \citep{Martig_2009, Genzel_2014}. This picture directly links quenching to morphological evolution, consistent with evidence that the likelihood of quiescence correlates not only with stellar mass but also with the bulge-to-total mass ratio (B/T) \citep{Bluck_2014, Omand_2014, Lang_2014, Shuntov_2025arXiv}. 

These two mechanisms may even act jointly rather than independently, considering the tight link between bulge growth and AGN activity at high redshift \citep{Daddi_2007b, ForsterSchreiber_2014, Rosario_2015, Mancini_2015}, and the well established bulge--supermassive black hole (SMBH) mass correlation observed in the local Universe \citep[][]{Magorrian_1998, Ferrarese_Merritt_2000, Kormendy_Ho_2013}.

\subsection{Mass-Quenching: Observational Constraints}
The first massive quiescent galaxies ($\log(M_*/M_{\odot}) \geq 10.5$) were already in place by $z \simeq 3$–4 \citep{DEugenio_2021, Carnall_2024, Genin_2025} but mass quenching reached its highest efficiency at cosmic noon ($1 \lesssim z \lesssim 3$), when both star formation and AGN activity reached their peak \citep{Madau_2014}, and the number density of massive quiescent systems increased by nearly two orders of magnitude while it continued to grow only mildly down to $z = 0$ \citep{Cimatti_2006,Ilbert_2013, Straatman_2014, Yang_2025arXiv}.

Several studies have focused on this redshift range, 
statistically confirming the strong connection between quenching and bulge growth in massive galaxies, through photometric bulge–disk decompositions of large samples of massive star-forming and quiescent galaxies up to $z\sim3$ \citep{Simard_2011, Bruce_2012, Lang_2014, Lange_2016, Dimauro_2018, Gentile_2025arXiv, Huertas-Company2025arXiv, Shuntov_2025arXiv}. In particular, there is a general consensus that the bulge-to-total mass ratio (B/T), or equivalently the S\'ersic index, increases with stellar mass and with decreasing specific SFR (sSFR) in star-forming galaxies along the MS \citep[e.g.][]{Wuyts_2011, Bruce_2012, Lang_2014, Tacchella_2015, Mancini_2015, Costantin_2021, Dimauro_2022, Genin_2025}.

At $z<1$, multi-band bulge–disk decompositions of large samples of both star-forming and quiescent galaxies have allowed stellar population ages and star formation histories (SFH) of bulges and disks to be inferred via Spectral Energy Distribution (SED) fitting \citep{Kennedy_2016, Margalef-Bentabol_2018, Mancini_2019, Costantin_2021, Costantin_2022, Gentile_2025arXiv}. These studies show that massive field galaxies quench inside-out, with bulges older and formed on much shorter timescales than disks \citep{Costantin_2021, Costantin_2022}. Two bulge formation pathways emerge: (i) an early population of compact, maximally old bulges quenched at $z>3$ and later rejuvenated through disk growth \citep{Mancini_2019}, and (ii) a younger, less compact population of bulges that quenched more gradually by $z\sim1$. 
Although photometric bulge-disk studies will soon extend to $z\sim3$ with current and upcoming high-resolution near-IR facilities (JWST/NIRCam, ELT/MICADO), ages derived from broad-band colours or SED fitting remain strongly degenerate with metallicity and dust reddening. Deep spectroscopy is therefore required to break these degeneracies \citep{Disney_1989, Worthey_1994}.

Spectroscopic “fossil record” analyses of stellar populations in quiescent galaxies provide key constraints on their quenching epochs and SFH timescales \citep{Gallazzi_2005, Renzini_2006, Gallazzi_2014, Carnall_2018, Wu_2018, Hamadouche_2022, Tacchella_2022, Ditrani_2025, Gallazzi_2025IarXiv, Gallazzi_2025IIarXiv, Mattolini_2025}. Thanks to increasingly powerful near-IR spectrographs, including NIRSpec/JWST, these diagnostics can now be pushed out to the cosmic noon \citep[e.g.][]{Belli_2019, Park_2023, Park_2024, Beverage_2025}. These studies likewise reveal a diversity of quenching timescales, with galaxies broadly following fast or slow pathways that yield quiescent remnants with distinct structural properties. Fast quenching produces compact, post-starburst remnants \citep{Dressler_1999} and becomes increasingly important at $z>2$. It is expected to be driven by violent mechanisms (e.g. major mergers or violent disk instabilities, VDI), capable of triggering an intense central burst of star formation and leading to galaxy compaction \citep{Tacchella_2016}. In contrast, slow quenching processes, thought to arise from the lack of replenishment of pristine gas from the circumgalactic medium \citep[starvation][]{Larson_1980}, lead to more extended quiescent systems and are observed to dominate at lower redshift \citep{Peng_2015}.  Together, these two modes help explain the strong size evolution of the quiescent population since $z\sim3$ \citep{Daddi_2005, VanDokkum_2008, Saracco_2011, vanderWel_2014, Gargiulo_2017, Genin_2025}. 

Moreover, JWST--NIRSpec–MSA has enabled spatially resolved measurements of stellar population gradients in quiescent galaxies at $1<z<2$ with unprecedented S/N \citep{Cheng_2025arXiv}. These results are consistent with an inside-out quenching scenario, in which galaxy cores appear systematically older yet more Mg-deficient than their outskirts, interpreted as earlier formation and more efficient quenching in the central regions, potentially driven by rapid gas removal that limited further chemical enrichment.

However, most of these observational efforts focus on already quenched systems, leaving the resolved properties and gradients of their massive star-forming and green-valley progenitors far less explored.

\subsection{Current Limitations to the Next Major Step}
\noindent A step forward would be to capture the massive star-forming progenitors of quiescent galaxies at cosmic noon in the act of quenching, using spatially resolved spectroscopy. 
In fact, high resolution and sensitivity are essential to separate the structural and kinematical components, such as bulges, disks clumps, bars, signs of merger, and outflows, and to identify the key signatures of the underlying physical processes \citep{ForsterSchreiber_2018}

In a pioneering study, \citet{Tacchella_2018} derived resolved SFR, M$_*$, and dust–attenuation maps for star-forming MS galaxies at $z\sim2$ by combining VLT--SINFONI H$\alpha$ Integral Field Spectrographs (IFS) with HST--WFC3 multi-band imaging. This provided the clearest evidence for inside-out quenching, on resolved scales, at cosmic noon. In fact, they found that all the most massive MS galaxies in their sample showed elevated sSFRs in their outskirts and suppressed, or quenched, star formation in their central regions, where a prominent bulge is in place.  
However, such analysis suffers from the age–dust-metallicity degeneracy inherent to broad-band–based stellar–population estimates, and further spectroscopic investigation are required. Current ground-based IFS still do not allow resolved stellar–population measurements capable to separate bulge- and disk- stellar populations at $z>1$. Their sensitivity is insufficient: prohibitively long exposure times would be required to reach the S/N$_{res} \gtrsim15$–20 needed to constrain stellar ages and metallicities to better than $\sim10-15\%$, especially for the low–surface-brightness exponential disks \citep[cf.][]{Mancini_2019}.

The VESPER IFS conceived for SHARP would, for the first time, provide the unique combination of medium resolution (R=3000), high sensitivity, sub-50 mas spatial resolution, and broad wavelength coverage in the optical rest-frame required to break key parameter degeneracies, and provide key insights on the physical mechanisms driving mass quenching. 

\section{Objectives}\label{sec:Obj}

\subsection{Probing Inside-Out Quenching with SHARP--VESPER} 
\noindent We propose to look for the signature of inside-out quenching at cosmic noon, by digging into the internal structure of galaxies that are plausibly direct progenitors of the most massive quiescent galaxies that populate the high-mass end of the red sequence at $1<z<3$. To this purpose, we aim to observe the most massive systems (log($M_\ast/M_\odot)\geq 11\pm0.1$) at $2.2<z<3.5$ that host a prominent bulge component but are still forming stars, either at a reduced rate compared to the MS (i.e. in the green valley) or still on the MS. Even MS galaxies in this mass regime are in fact expected to quench on short timescales ($\lesssim 1$ Gyr), as sustaining MS-level SFRs would otherwise lead to an excessive stellar-mass growth that is not observed.

For these galaxies, we have the ambitious goal of obtaining, in a single observation, both deep stellar–continuum spectra with a S/N$_{res}$ $>15$-20 for the (inner) bulge and (outer) disk components, and the spatial distribution of ionised gas traced by nebular emission lines. When complemented with spatially resolved photometry (e.g., from JWST, or ELT), this will allow us to constrain, on the one hand, the ages, mass distributions, chemical abundances, and SFHs of bulges and disks separately. On the other hand, it will enable us to map the spatially resolved SFR, characterise the gas kinematics, identify star-forming clumps, detect stellar- or AGN-driven outflows, or radial gas flows (e.g. feeding star-formation in the central region).

\subsection{Sample: Galaxies in the Act of Quenching} \label{sec:Sample}
The 0.54~deg$^2$ COSMOS-Web field \citep{Casey_2023} provides an ideal basis for our target selection, offering extensive multi-wavelength photometry, redshifts, and high-resolution NIRCam and MIRI imaging, along with structural and morphological information \citep{Shuntov_2025arXiv, Huertas-Company2025arXiv}.

From the COSMOS-Web master catalogue we select the H-band brightest (mag\_auto$_{\rm F150W}<23.5$) and most massive (log(M$\ast$/M$_\odot$)$\geq11\pm0.1$) galaxies at $2.2<z<3.5$, to keep SHARP--VESPER integration times manageable. This redshift interval ensures full coverage of the optical rest-frame (from 3750-7500~\AA\ to 2670-5330~\AA\ across the range) with SHARP--VESPER broad 1.2–2.4~$\mu$m spectral range. 
This wavelength coverage is optimal for mitigating the age–metallicity degeneracy \citep[][and references therein]{Gallazzi_2005, vanderWel_2014}, as it includes all key diagnostic features sensitive to stellar age, metallicity, and $\alpha$-element abundances, such as the Balmer lines, Ca~H+K, the 4000~\AA\ break, and prominent CN, Fe, Ca, and Mg absorption features.

\begin{figure*}
\centering
\includegraphics[width=0.9\textwidth]{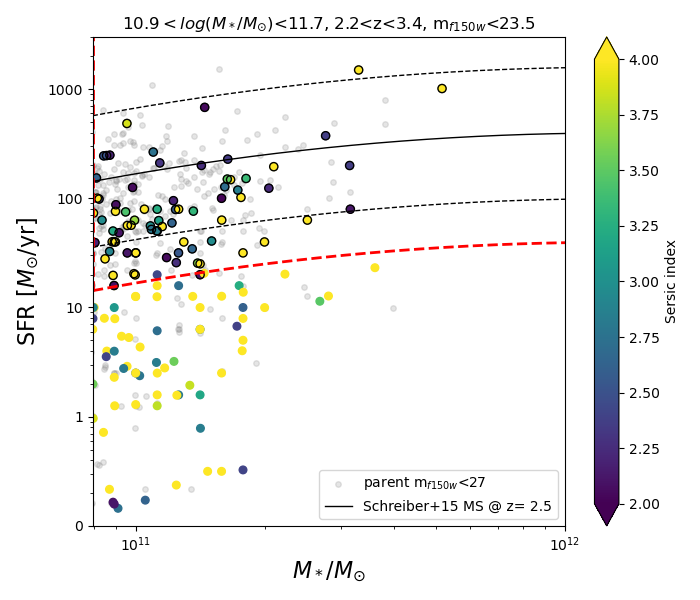}\\
\includegraphics[width=.13\textwidth]{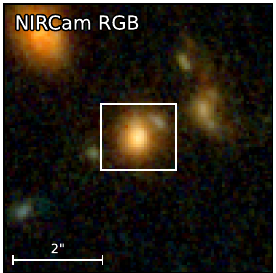}
\includegraphics[width=.13\textwidth]{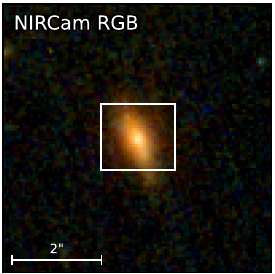}
\includegraphics[width=.13\textwidth]{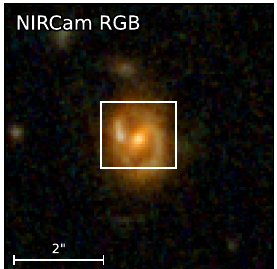}
\includegraphics[width=.13\textwidth]{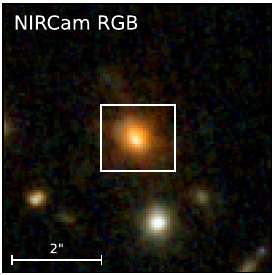}
\includegraphics[width=.13\textwidth]{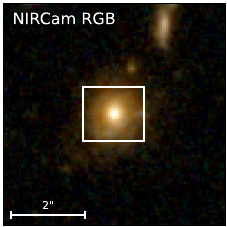}
\includegraphics[width=.13\textwidth]{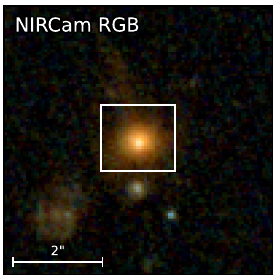}\\
\includegraphics[width=.13\textwidth]{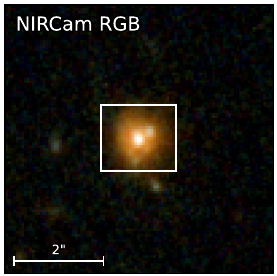}
\includegraphics[width=.13\textwidth]{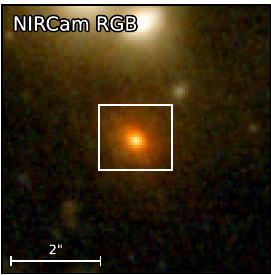}
\includegraphics[width=.13\textwidth]{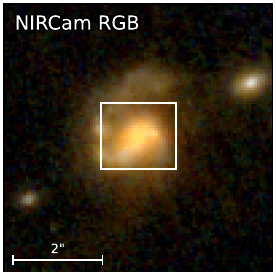}
\includegraphics[width=.13\textwidth]{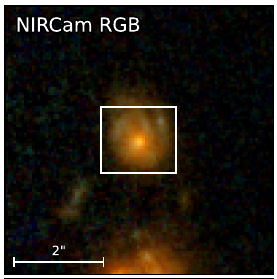}
\includegraphics[width=.13\textwidth]{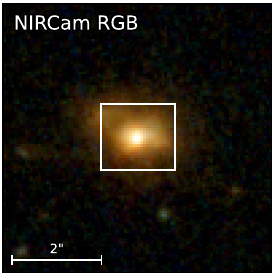}
\includegraphics[width=.13\textwidth]{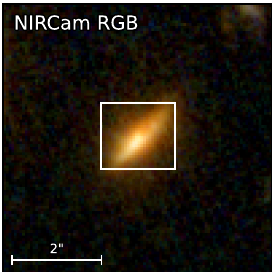}\\
\caption{\textbf{Top:} Position of our target galaxies at $2.2<z_{\rm phot}<3.5$ on the $M_\ast$–SFR plane, compared to the star-forming MS, and its $2.5\sigma$ scatter, from \citet{Schreiber_2015} at the median redshift $z\sim2.5$ (black solid and dashed lines). Grey filled circles show galaxies from the COSMOS-Web parent catalogue in the same redshift range. Our targets (black-edged circles, colour-coded by S\'ersic index) comprise the brightest galaxies in this mass range with prominent bulges (S\'ersic $n>2$) and specific star formation rates consistent with the MS or up to a factor of $\sim$10 below it. The red dashed line indicates this selection limit in terms of distance from the MS, defined as ${\rm sSFR(MS)}/{\rm sSFR} \leq 10$. \textbf{Bottom: } Gallery of NIRCam RGB images illustrating the typical appearance of our targets. The white rectangles indicate the $1\farcs7\times 1\farcs5$ SHARP--VESPER IFU field of view.}
\label{FIG:1}
\end{figure*}

As mentioned above, our target selection further requires:
(1) the presence of a prominent bulge ($B/T>0.3$ or S\'ersic index $n>2$; cf. \citealt{Mancini_2015}, \citealt{Shuntov_2025arXiv}), to select galaxies that are candidates for inside-out quenching; and 
(2) a lower limit in sSFR that includes galaxies on the MS or down to a factor of 10 ($\sim6.5\sigma$) below it, i.e. within the green valley.
As shown in Section~\ref{sec:TechJust}, this lower limit on sSFR is chosen based on the limiting line flux required to ensure the detection of nebular emission lines ([OII], [OIII], H$\beta$, H$\alpha$, and [NII]), with a S/N$_{res}$>5 on spatially resolved sub-kpc scales in a reasonable amount of time. 
In Fig.~\ref{FIG:1} (top panel) we show the position of our target galaxies (colour–coded by S\'ersic index) and of the COSMOS-Web parent sample at the same redshift (grey points) relative to the average MS at $z\sim2.5$.
The red dashed region marks our sample selection, illustrating the applied limit on sSFR in terms of distance from the MS (sSFR(MS)/sSFR~$\leq 10$).

In the bottom panel of Fig.~\ref{FIG:1} we show a few example RGB images of galaxies in the sample \citep[from the COSMOS-Web archive,][]{Franco_2025arXiv} with the SHARP--VESPER field of view overlaid. The sample  spans the following range (median) properties:
$R_{\rm e} = 0.07$–$0.4$ ($0.26\pm0.16$)~arcsec,
$n = 2$–$8$ ($3.0\pm1.4$),
and mag\_auto$_{\rm F150W} = 21$–$23.5$ ($22.6\pm0.44$).
As can be seen, the selected population covers a broad range of sizes, since our sample naturally includes both compact and extended systems. 

We will ensure comprehensive coverage of the relevant parameter space by sampling both the offset from the star-forming main sequence (sSFR(MS)/sSFR), from MS to green-valley systems, and the galaxy size, as traced by the S\'ersic half-light radius ($R_{\rm e}$).
To this end, the sample will be divided into four equally populated bins in distance from the MS.

Within each bin, we will select a comparable number of galaxies (typically 5-7), for a total of 20-30 targets, while preserving a balanced mix of compact ($R_{\rm e}<180$~mas), intermediate ($180<R_{\rm e}<250$~mas), and extended ($R_{\rm e}>250$~mas) systems, relative to the parent star-forming population \citep[see][]{Genin_2025}.
This strategy ensures that, even if on a relatively small sample, we probe different structural states and levels of star-formation suppression. This will allow us to test whether bulge growth and quenching proceed along distinct evolutionary pathways as a function of compactness and MS offset. It is worth to note that, capturing the extremely compact galaxies (with $R_{\rm e}\sim0.1$~arcsec) is only feasible thanks to the sub-50~mas spatial sampling provided by SHARP--VESPER.
     
\subsection{Expected Outcomes}
\noindent The proposed observations with SHARP--VESPER will enable several key analyses, as summarised below. The combination of all these diagnostics will allow us to directly test the interconnection and relative  timescales of the processes linked to bulge growth and quenching, and to assess how these mechanisms operate in the most massive galaxies at cosmic noon. The main expected outcomes are:

\noindent \textbf{$\bullet$~Constraining bulge and disk mass assembly history.}
With the planned integration time, we will achieve S/N$_{\rm res} \gtrsim 15$–20 on the stellar continuum for both bulge and disk components, by integrating the signal over spaxels in the inner (r < R$_{\rm e,B}$) and outer (R$_{\rm e,B}$ < r < R$_{\rm e,D}$) regions, where $R_{\rm e,B}$ and $R_{\rm e,D}$ 
denote the bulge and disk half-light radii, respectively. For each component, medium-resolution optical rest-frame spectra (R~$\sim$~3000) will allow independent measurements of ages, metallicities, $\alpha$-element abundances, stellar masses, dust attenuation, and star-formation histories, through Lick indices analysis, and full spectral fitting. Such bulge–disk spectroscopic decomposition breaks the degeneracies that affect spatially integrated spectra and enables a direct reconstruction of the mass assembly history of the two structural components. This analysis may be further supported by photometric bulge–disk decomposition based on high-resolution ELT imaging.

\noindent \textbf{$\bullet$~Constraining the gas distribution, metallicity, and kinematics.}
Our sSFR selection ensures the detection of nebular lines with fluxes $\gtrsim 1.5\times10^{-17}$ erg s$^{-1}$ cm$^{-2}$, enabling measurements of [O II], H$\beta$, [O III], and H$\alpha$+[N II] (the latter up to $z\sim2.6$). With the planned integration, we expect S/N$_{res}$>5 per spatial bin, on sub-kpc scales (or kpc scales with 3$\times$3 spatial rebinning in the lowest surface brightness regions), allowing the study of the ionised-gas distribution across the disks, and of gas kinematics, as well as the detection of star-forming clumps \citep{Genzel_2011}. 
H$\alpha$ and/or [O II], corrected for dust, will provide spatially resolved SFRs \citep[cf.,][]{Kennicutt_1998}, while multiple detected lines will allow us to constrain gas-phase metallicity \citep{Maiolino_2019}, excitation conditions, and outflow signatures (see next item). When possible, spatially resolved dust attenuation in the nebular regions will be derived from the Balmer decrement \citep[H$\alpha$/H$\beta$][]{Osterbrock_1989} and compared to the stellar continuum attenuation inferred from spectro-photometric SED fitting.

\noindent \textbf{$\bullet$~Feedback Signatures and the Origins of Outflows.}
A significant incidence of AGN is expected in such massive systems. Spatially resolved emission-line maps will allow us to search for feedback-driven features, revealed by broadened (FWHM>400-500 km/s) or asymmetric line profiles. When detected, the combination of line widths and diagnostic ratios (e.g.\ [OIII]/H$\beta$, [NII]/H$\alpha$) will enable us to distinguish AGN-driven outflows from those powered by star formation \citep{ForsterSchreiber_2014, ForsterSchreiber_2019}.

\noindent \textbf{$\bullet$~Testing Bulge Formation Pathways.}
Dynamical models predict that classical bulges (old, dispersion-dominated spheroids, with high $\alpha$/Fe ratios) form rapidly through centrally concentrated starbursts, driven by early collapse \citep{Eggen_1962}, the inward migration and coalescence of massive clumps ($\lesssim$1 Gyr; \citealt{Noguchi_1999, Immeli_2004, Genzel_2008, Dekel_2009b, Dekel_2014}), or mergers \citep{Scannapieco_2003}. This scenario is supported by high-resolution ALMA observations, which reveal spheroidal morphologies of the compact dust cores in luminous sub-millimetre galaxies at $1.5<z<3$, consistent with in-situ bulge formation \citep{Tan_2024}.
In contrast, bluer and rotation-dominated pseudo-bulges are expected to form on longer timescales via processes such as bar-driven inflows \citep{Combes_Sanders_1981, Debattista_2006}. With our data we can test these scenarios by measuring bulge ages and $\alpha$-enhancement (short, bursty vs. extended formation) and by identifying structural or kinematic signatures of clumps, bars, or merger-driven disturbances.

\section{Technical Justification}\label{sec:TechJust}

We used the official SHARP--VESPER Exposure Time Calculator (ETC, version~0.6\footnote{\href{https://sharp.lambrate.inaf.it/}{https://sharp.lambrate.inaf.it/}}) to estimate the integration times required for our observations.
As detailed in Section~\ref{sec:Obj}, the first and distinctive goal of this program is to obtain a S/N$_{res}>15$–20 on the spectral continuum for both the bulge and disk components of our $z=2$–3 galaxies, enabling robust measurements of stellar ages, stellar mass, chemical abundances, and SFHs. 

The second goal is to detect the nebular emission lines tracing the ionised gas distribution, with a sensitivity sufficient to reach S/N$_{res}$>5 per spaxel, allowing us to study gas kinematics and to identify possible outflow signatures.

In the following, we present the integration-time estimates required to achieve these two main objectives.
\subsection{Bulge and Disk Spectral Continuumm}
To ensure that our required S/N$_{res}$ is reached in both the bulge and disk components, we simulated them separately using the SHARP--VESPER ETC with default configuration (Airmass=1.5 ZD=48 deg, Spectral binning [pixels]=4,  AO-mode = MCAO) and adopting the extended morphology option. The structural parameters for each component were taken from the COSMOS-Web bulge–disk morphological measurements \citep{Genin_2025}, which assume an exponential S\'ersic surface-brightness profiles for the disk and a \citet[][$n=4$]{deVaucouleurs_1953} profile for the bulge. As detailed in Section~\ref{sec:Sample}, our bright, massive galaxies at $2.2\leq z\leq 3.5$ host, by selection, a significant bulge component. For the following computations, we adopt a median H-band bulge-to-total ratio of B/T$\sim0.4$ (from mag\_auto$_{F150W}$) to derive the corresponding bulge and disk H-band magnitudes. 
Because achieving high continuum S/N$_{\rm res}$ in extended exponential disks requires substantially longer integration times than for compact bulges, we set the total exposure time by imposing the S/N$_{\rm res}$ requirement on the disk component. Specifically, this is evaluated in the annulus between the bulge and disk half-light radii (R${\rm e,B}<r<$R${\rm e,D}$), where the bulge contribution is negligible, thereby ensuring a cleaner bulge–disk separation in the central regions. 
 
We simulated three representative cases drawn from our sample:
(i) a galaxy with average properties (Avg),
(ii) a less optimistic scenario with a faint, extended disk (Faint-E), and
(iii) the most favourable case of a bright, compact source (Bright-C).
For each case, we adopted the Simple Stellar Population (SSP) templates available in the SHARP--VESPER ETC \citep{Bruzual_2003}, assuming ages of 2 Gyr for the bulge and 0.1 Gyr for the disk component. 
The results are summarised in Table~\ref{tbl:1}, which lists the main input parameters used in the ETC, followed by the S/N$_{\rm res, D}$ requirement in the outer disk, the resulting S/N$_{\rm res, B}$ in the inner bulge, and the total exposure time needed (in hours). In summary, the required exposure time spans from $\sim$5 hours in the bright compact cases to $\sim$30 hours in the faint extended-disk systems, with a typical value of $\sim$15 hours for most (average) sources.

\begin{table}[width=.95\linewidth,cols=9,pos=h]
\footnotesize
\caption{Integration times from SHARP--VESPER ETC for three representative cases of galaxies in our sample.}\label{tbl:1}
\begin{tabular*}{\tblwidth}{@{} LRRRRRRRRR@{} }
\toprule
Target & z& H$_{Tot}$ & H$_{B}$ & H$_{D}$ &R$_{e,B}$ & R$_{e,D}$ & S/N$_{res,D}$ & S/N$_{res,B}$ & Texp \\  
      &   & [AB] & [AB] &[AB] &[mas] &[mas] &  & &[hr]\\
\midrule
Avg   &2.5&22.6&23.6&23.1&100&200&20&30 &15\\
Faint-E  & 3.0& 23.4&24.4&23.9&100&350&15&23 &30\\
Bright-C  &2.3 &22.0 &23.0&22.5&50&150&28&40 &5\\
\end{tabular*}
\end{table}

\subsection{Nebular Emission Lines}
For nebular emission lines, we estimated the expected [O II] and H$\alpha$ fluxes using the SFR and the stellar-continuum colour excess $E(B-V)$ provided by the COSMOS-Web catalogue. Line luminosities are derived using the SFR – luminosity calibrations from \citet{Kennicutt_1998}, and corrected for dust attenuation adopting the \citet{Calzetti_2000} extinction law, assuming (conservatively) that nebular lines are attenuated by an additional factor of $\sim$1.7 relative to the stellar continuum. On average, our targets are therefore expected to exhibit line fluxes of F[O II] $\sim 1.5\times 10^{-17}$ and FH$\alpha$ $\sim 3\times 10^{-17}$ [erg s$^{-1}$ cm$^{-2}$].

Although our sample spans a wide range of SFR ($\sim$ 20–1500 M$_{\odot}$,yr$^{-1}$) and in H$\alpha$ flux ($1.5\times 10^{-18}-2\times10^{-16}$[erg s$^{-1}$ cm$^{-2}$]), the increasing dust attenuation at higher SFRs keeps the typical observed fluxes close to these average values.

Also in this case we used the SHARP--VESPER ETC with default configuration (Airmass=1.5 ZD=48 deg, Spectral binning [pixels]=4, AO-mode = MCAO) providing as input our average F[OII] and FH$\alpha$ line fluxes, assuming a FWHM = 150-200 km/s \citep[][]{ForsterSchreiber_2018}, and the reference wavelengths of the expected [OII], and H$\alpha$ emission lines at each given redshift.

For the three representative cases listed in Table~\ref{tbl:1}, we computed the S/N obtained for nebular emission lines using the exposure times derived from the continuum requirement, adopting the same Extended morphology parameters used for the stellar disks. In all cases, we verified that the integration times needed to reach the target continuum S/N$_{res}$ are sufficient to recover spatially resolved H$\alpha$ and/or [O II] maps with S/N$_{res}$> 5 per spatial bin (on sub-kpc scales, or kpc scales with 3$\times$3 spatial rebinning in the lowest surface brightness regions), out to at least $\sim$1–1.5 disk half-light radii ($R_{e,D}$).

\subsection{Exploiting SHARP--VESPER Multiplexing} 
The extremely massive, bright galaxies presented in this study are exceptionally rare at $z=2$–3, and can therefore be observed only one per SHARP--VESPER pointing. Their scarcity, however, also underlines the need for careful selection and dedicated follow-up. 
In a potential SHARP--VESPER large survey, these systems could be prioritised as primary targets, while the remaining 11 IFUs are allocated to more common galaxies requiring comparable exposure times ($\sim$15–20 hours). 

A particularly effective complementary program is that of Vietri et al.\ in this Science Book, based on the same COSMOS-Web catalogue in the COSMOS field, targeting galaxies at $1.5 < z < 2.5$ over a wide mass range ($\log(M_*/\rm M_{\odot}) = 8$–11.5) and with SFR $\geq 1\, \rm M_{\odot}\,yr^{-1}$. The high surface density of their sample  
ensures $\sim10$–15 galaxies per SHARP--VESPER FoV, with required exposure times of $\sim20$ hours, which are fully compatible with, and even exceed, those needed to reach S/N $\sim 20$ in the disk component of our targets (see Table~\ref{tbl:1}).

This strategy implies $\sim 20$–30 SHARP--VESPER pointings, by allocating one IFU to each of our rare systems and filling the remaining IFUs with Vietri et al.\ targets. In this way, we can observe $\sim 20$–30 of our key galaxies within a total sample of $\sim 200$–300 objects, over a total investment of $\sim 300$–600 hours, fully exploiting SHARP--VESPER's multiplexing capabilities.

\section{Comparison with Other Facilities}
We stress that SHARP--VESPER on the ELT, in combination with the Multi-Conjugate Adaptive Optics (MCAO) system, would represent an unrivalled instrument for carrying out the proposed observations.

Thanks to its high sensitivity and exceptional spatial resolution (31~mas, more than a factor of three sharper than JWST $\sim$100~mas), SHARP--VESPER will for the first time enable a spectroscopic characterisation of the stellar populations within the central kiloparsec of our $z = 2.2$–$3.5$ targets, with unprecedented spatial sampling. This is essential for our massive star-forming galaxies, as this region hosts ultra-dense bulges (with sub-kpc half-light radii) and may retain the signatures of recent quenching, including post-starburst features \citep{Wild_2009} or spectral imprints of outflows \citep{Maltby_2019}. This becomes even more critical for the most compact systems in our sample (e.g., Bright-C in Table~\ref{tbl:1}).
For the ionised gas component, SHARP--VESPER enhanced angular resolution will also allow us to resolve star-forming clumps and to identify outflows or merger-driven perturbations on sub-kiloparsec scales, for the first time at these redshifts \citep[cf., ][]{ForsterSchreiber_2018}. 

Among next-generation ELT IFS facilities, HARMONI \citep{Thatte_2024} will offer spatial sampling comparable to $-$and optionally even finer than$-$ SHARP--VESPER, together with similar sensitivity and near-IR wavelength coverage. However, in a large program survey perspective, SHARP--VESPER offers a clear efficiency advantage thanks to its multiplexing capability of 12 Integral Field Units (IFU). 

\section*{Acknowledgments}
\noindent The SHARP team acknowledges support by Bando Ricerca Fondamentale INAF 2022, Techno-Grant "SHARP" - 1. 05. 12. 02. 01 and Bando Ricerca Fondamentale INAF 2024, Large-Grant "SHARP" - 1. 05. 24. 01. 01.

\printcredits


\bibliographystyle{cas-model2-names}

\bibliography{bd}

@ARTICLE{deVaucouleurs_1953,
       author = {{de Vaucouleurs}, G.},
        title = "{On the distribution of mass and luminosity in elliptical galaxies}",
      journal = {\mnras},
         year = 1953,
        month = jan,
       volume = {113},
        pages = {134},
          doi = {10.1093/mnras/113.2.134},
       adsurl = {https://ui.adsabs.harvard.edu/abs/1953MNRAS.113..134D}
}

@article{Eggen_1962,
  author    = {Eggen, O. J. and Lynden--Bell, D. and Sandage, R. A.},
  title     = {Evidence from the Motions of Old Stars that the Galaxy Collapsed},
  journal   = {Astrophysical Journal},
  volume    = {136},
  pages     = {748--766},
  year      = {1962},
  adsurl    = {https://ui.adsabs.harvard.edu/abs/1962ApJ...136..748E}
}

@BOOK{Sersic_1968,
   author = {{Sersic}, J.~L.},
    title = "{Atlas de galaxias australes}",
publisher = {Cordoba, Argentina: Observatorio Astronomico, 1968},
     year = 1968,
   adsurl = {http://adsabs.harvard.edu/abs/1968adga.book.....S}
}

@ARTICLE{Gunn_1972,
       author = {{Gunn}, James E. and {Gott}, III, J. Richard},
        title = "{On the Infall of Matter Into Clusters of Galaxies and Some Effects on Their Evolution}",
      journal = {\apj},
         year = 1972,
        month = aug,
       volume = {176},
        pages = {1},
          doi = {10.1086/151605},
       adsurl = {https://ui.adsabs.harvard.edu/abs/1972ApJ...176....1G}
}

@ARTICLE{Combes_Sanders_1981,
       author = {{Combes}, F. and {Sanders}, R.~H.},
        title = "{Formation and properties of persisting stellar bars.}",
      journal = {\aap},
         year = 1981,
        month = mar,
       volume = {96},
        pages = {164-173},
       adsurl = {https://ui.adsabs.harvard.edu/abs/1981A&A....96..164C}
}

@ARTICLE{Larson_1980,
       author = {{Larson}, R.~B. and {Tinsley}, B.~M. and {Caldwell}, C.~N.},
        title = "{The evolution of disk galaxies and the origin of S0 galaxies}",
      journal = {\apj},
         year = 1980,
        month = may,
       volume = {237},
        pages = {692-707},
          doi = {10.1086/157917},
       adsurl = {https://ui.adsabs.harvard.edu/abs/1980ApJ...237..692L}
}

@article{Disney_1989,
  author    = {Disney, M. and Davies, J. and Phillipps, S.},
  title     = {Are galaxy discs optically thick?},
  journal   = {Monthly Notices of the Royal Astronomical Society},
  volume    = {239},
  pages     = {939--976},
  year      = {1989},
  doi       = {10.1093/mnras/239.3.939},
  adsurl    = {https://ui.adsabs.harvard.edu/abs/1989MNRAS.239..939D}
}

@BOOK{Osterbrock_1989,
       author = {{Osterbrock}, Donald E.},
        title = "{Astrophysics of gaseous nebulae and active galactic nuclei}",
         year = 1989,
       adsurl = {https://ui.adsabs.harvard.edu/abs/1989agna.book.....O}
}

@ARTICLE{Worthey_1994,
       author = {{Worthey}, Guy},
        title = "{Comprehensive Stellar Population Models and the Disentanglement of Age and Metallicity Effects}",
      journal = {\apjs},
         year = 1994,
        month = nov,
       volume = {95},
        pages = {107},
          doi = {10.1086/192096},
       adsurl = {https://ui.adsabs.harvard.edu/abs/1994ApJS...95..107W}
}

@ARTICLE{Moore_1996,
       author = {{Moore}, Ben and {Katz}, Neal and {Lake}, George and {Dressler}, Alan and {Oemler}, Augustus},
        title = "{Galaxy harassment and the evolution of clusters of galaxies}",
      journal = {\nat},
         year = 1996,
        month = feb,
       volume = {379},
       number = {6566},
        pages = {613-616},
          doi = {10.1038/379613a0},
archivePrefix = {arXiv},
       eprint = {astro-ph/9510034},
 primaryClass = {astro-ph},
       adsurl = {https://ui.adsabs.harvard.edu/abs/1996Natur.379..613M}
}

@ARTICLE{Kennicutt_1998,
   author = {{Kennicutt}, Jr., R.~C.},
    title = "{Star Formation in Galaxies Along the Hubble Sequence}",
  journal = {\araa},
   eprint = {arXiv:astro-ph/9807187},
     year = 1998,
   volume = 36,
    pages = {189-232},
      doi = {10.1146/annurev.astro.36.1.189},
   adsurl = {http://adsabs.harvard.edu/abs/1998ARA%26A..36..189K}
}

@article{Magorrian_1998,
  author       = {Magorrian, J. and Tremaine, S. and Richstone, D. and Bender, R. and Bower, G. and Dressler, A. and Faber, S.~M. and Gebhardt, K. and Green, R. and Grillmair, C. and Kormendy, J. and Lauer, T.~R.},
  title        = {The Demography of Massive Dark Objects in Galaxy Centers},
  journal      = {Astronomical Journal},
  volume       = {115},
  pages        = {2285--2305},
  year         = {1998},
  doi          = {10.1086/300353},
  adsurl       = {https://ui.adsabs.harvard.edu/abs/1998AJ....115.2285M}
}

@ARTICLE{Dressler_1999,
       author = {{Dressler}, A. and {Smail}, I. and {Poggianti}, B.~M. and {Butcher}, H. and 
                 {Couch}, W.~J. and {Ellis}, R.~S. and {Oemler}, A.},
        title = "{The Morphologies of Distant Cluster Galaxies}",
      journal = {\apjs},
         year = 1999,
        month = jun,
       volume = {122},
        pages = {51--80},
          doi = {10.1086/313213},
       adsurl = {https://ui.adsabs.harvard.edu/abs/1999ApJS..122...51D}
}

@ARTICLE{Noguchi_1999,
       author = {{Noguchi}, Masafumi},
        title = "{Early Evolution of Disk Galaxies: Formation of Bulges in Clumpy Young Galactic Disks}",
      journal = {\apj},
         year = 1999,
        month = mar,
       volume = {514},
       number = {1},
        pages = {77-95},
          doi = {10.1086/306932},
archivePrefix = {arXiv},
       eprint = {astro-ph/9806355},
 primaryClass = {astro-ph},
       adsurl = {https://ui.adsabs.harvard.edu/abs/1999ApJ...514...77N}
}

@ARTICLE{Calzetti_2000,
   author = {{Calzetti}, D. and {Armus}, L. and {Bohlin}, R.~C. and {Kinney}, A.~L. and 
	{Koornneef}, J. and {Storchi-Bergmann}, T.},
    title = "{The Dust Content and Opacity of Actively Star-forming Galaxies}",
  journal = {\apj},
   eprint = {arXiv:astro-ph/9911459},
     year = 2000,
    month = apr,
   volume = 533,
    pages = {682-695},
      doi = {10.1086/308692},
   adsurl = {http://adsabs.harvard.edu/abs/2000ApJ...533..682C}
}

@article{Ferrarese_Merritt_2000,
  author    = {Ferrarese, L. and Merritt, D.},
  title     = {A Fundamental Relation Between Supermassive Black Holes and Their Host Galaxies},
  journal   = {The Astrophysical Journal Letters},
  volume    = {539},
  pages     = {L9--L12},
  year      = {2000},
  doi       = {10.1086/312838},
  adsurl    = {https://ui.adsabs.harvard.edu/abs/2000ApJ...539L...9F}
}

@ARTICLE{Trujillo_2001,
       author = {{Trujillo}, I. and {Graham}, Alister W. and {Caon}, N.},
        title = "{On the estimation of galaxy structural parameters: the S{\'e}rsic model}",
      journal = {\mnras},
         year = 2001,
        month = sep,
       volume = {326},
       number = {3},
        pages = {869-876},
          doi = {10.1046/j.1365-8711.2001.04471.x},
archivePrefix = {arXiv},
       eprint = {astro-ph/0102393},
 primaryClass = {astro-ph},
       adsurl = {https://ui.adsabs.harvard.edu/abs/2001MNRAS.326..869T}
}

@ARTICLE{Birnboim_2003,
       author = {{Birnboim}, Yuval and {Dekel}, Avishai},
        title = "{Virial shocks in galactic haloes?}",
      journal = {\mnras},
         year = 2003,
        month = oct,
       volume = {345},
       number = {1},
        pages = {349-364},
          doi = {10.1046/j.1365-8711.2003.06955.x},
archivePrefix = {arXiv},
       eprint = {astro-ph/0302161},
 primaryClass = {astro-ph},
       adsurl = {https://ui.adsabs.harvard.edu/abs/2003MNRAS.345..349B}
}

@ARTICLE{Bruzual_2003,
   author = {{Bruzual}, G. and {Charlot}, S.},
    title = "{Stellar population synthesis at the resolution of 2003}",
  journal = {\mnras},
   eprint = {arXiv:astro-ph/0309134},
     year = 2003,
    month = oct,
   volume = 344,
    pages = {1000-1028},
      doi = {10.1046/j.1365-8711.2003.06897.x},
   adsurl = {http://adsabs.harvard.edu/abs/2003MNRAS.344.1000B}
}

@ARTICLE{Scannapieco_2003,
       author = {{Scannapieco}, C. and {Tissera}, P.~B.},
        title = "{The effects of mergers on the formation of disc-bulge systems in hierarchical clustering scenarios}",
      journal = {\mnras},
         year = 2003,
        month = feb,
       volume = {338},
       number = {4},
        pages = {880-890},
          doi = {10.1046/j.1365-8711.2003.06133.x},
archivePrefix = {arXiv},
       eprint = {astro-ph/0208538},
 primaryClass = {astro-ph},
       adsurl = {https://ui.adsabs.harvard.edu/abs/2003MNRAS.338..880S}
}

@ARTICLE{Binney_2004,
       author = {{Binney}, James},
        title = "{On the origin of the galaxy luminosity function}",
      journal = {\mnras},
         year = 2004,
        month = feb,
       volume = {347},
       number = {4},
        pages = {1093-1096},
          doi = {10.1111/j.1365-2966.2004.07277.x},
archivePrefix = {arXiv},
       eprint = {astro-ph/0308172},
 primaryClass = {astro-ph},
       adsurl = {https://ui.adsabs.harvard.edu/abs/2004MNRAS.347.1093B}
}

@ARTICLE{Granato_2004,
       author = {{Granato}, Gian Luigi and {De Zotti}, Gianfranco and {Silva}, Laura and {Bressan}, Alessandro and {Danese}, Luigi},
        title = "{A Physical Model for the Coevolution of QSOs and Their Spheroidal Hosts}",
      journal = {\apj},
         year = 2004,
        month = jan,
       volume = {600},
       number = {2},
        pages = {580-594},
          doi = {10.1086/379875},
archivePrefix = {arXiv},
       eprint = {astro-ph/0307202},
 primaryClass = {astro-ph},
       adsurl = {https://ui.adsabs.harvard.edu/abs/2004ApJ...600..580G}
}

@ARTICLE{Immeli_2004,
       author = {{Immeli}, A. and {Samland}, M. and {Gerhard}, O. and {Westera}, P.},
        title = "{Gas physics, disk fragmentation,  and bulge formation in young galaxies}",
      journal = {\aap},
         year = 2004,
        month = jan,
       volume = {413},
        pages = {547-561},
          doi = {10.1051/0004-6361:20034282},
archivePrefix = {arXiv},
       eprint = {astro-ph/0312139},
 primaryClass = {astro-ph},
       adsurl = {https://ui.adsabs.harvard.edu/abs/2004A&A...413..547I}
}

@ARTICLE{Gallazzi_2005,
   author = {{Gallazzi}, A. and {Charlot}, S. and {Brinchmann}, J. and {White}, S.~D.~M. and 
	{Tremonti}, C.~A.},
    title = "{The ages and metallicities of galaxies in the local universe}",
  journal = {\mnras},
   eprint = {astro-ph/0506539},
     year = 2005,
    month = sep,
   volume = 362,
    pages = {41-58},
      doi = {10.1111/j.1365-2966.2005.09321.x},
   adsurl = {http://adsabs.harvard.edu/abs/2005MNRAS.362...41G}
}

@ARTICLE{Keres_2005,
       author = {{Kere{\v{s}}}, Du{\v{s}}an and {Katz}, Neal and {Weinberg}, David H. and {Dav{\'e}}, Romeel},
        title = "{How do galaxies get their gas?}",
      journal = {\mnras},
         year = 2005,
        month = oct,
       volume = {363},
       number = {1},
        pages = {2-28},
          doi = {10.1111/j.1365-2966.2005.09451.x},
archivePrefix = {arXiv},
       eprint = {astro-ph/0407095},
 primaryClass = {astro-ph},
       adsurl = {https://ui.adsabs.harvard.edu/abs/2005MNRAS.363....2K}
}

@ARTICLE{Reddy_2005,
       author = {{Reddy}, Naveen A. and {Erb}, Dawn K. and {Steidel}, Charles C. and {Shapley}, Alice E. and {Adelberger}, Kurt L. and {Pettini}, Max},
        title = "{A Census of Optical and Near-Infrared Selected Star-forming and Passively Evolving Galaxies at Redshift z \raisebox{-0.5ex}\textasciitilde 2}",
      journal = {\apj},
         year = 2005,
        month = nov,
       volume = {633},
       number = {2},
        pages = {748-767},
          doi = {10.1086/444588},
archivePrefix = {arXiv},
       eprint = {astro-ph/0507264},
 primaryClass = {astro-ph},
       adsurl = {https://ui.adsabs.harvard.edu/abs/2005ApJ...633..748R}
}

@ARTICLE{Daddi_2005,
   author = {{Daddi}, E. and {Renzini}, A. and {Pirzkal}, N. and {Cimatti}, A. and 
	{Malhotra}, S. and {Stiavelli}, M. and {Xu}, C. and {Pasquali}, A. and 
	{Rhoads}, J.~E. and {Brusa}, M. and {di Serego Alighieri}, S. and 
	{Ferguson}, H.~C. and {Koekemoer}, A.~M. and {Moustakas}, L.~A. and 
	{Panagia}, N. and {Windhorst}, R.~A.},
    title = "{Passively Evolving Early-Type Galaxies at 1.4 {\lt}\~{} z {\lt}\~{} 2.5 in the Hubble Ultra Deep Field}",
  journal = {\apj},
   eprint = {arXiv:astro-ph/0503102},
     year = 2005,
    month = jun,
   volume = 626,
    pages = {680-697},
      doi = {10.1086/430104},
   adsurl = {http://adsabs.harvard.edu/abs/2005ApJ...626..680D}
}

@ARTICLE{Baldry_2006,
       author = {{Baldry}, I.~K. and {Balogh}, M.~L. and {Bower}, R.~G. and {Glazebrook}, K. and {Nichol}, R.~C. and {Bamford}, S.~P. and {Budavari}, T.},
        title = "{Galaxy bimodality versus stellar mass and environment}",
      journal = {\mnras},
         year = 2006,
        month = dec,
       volume = {373},
       number = {2},
        pages = {469-483},
          doi = {10.1111/j.1365-2966.2006.11081.x},
archivePrefix = {arXiv},
       eprint = {astro-ph/0607648},
 primaryClass = {astro-ph},
       adsurl = {https://ui.adsabs.harvard.edu/abs/2006MNRAS.373..469B}
}

@ARTICLE{Cimatti_2006,
       author = {{Cimatti}, A. and {Daddi}, E. and {Renzini}, A.},
        title = "{Mass downsizing and ``top-down'' assembly of early-type galaxies}",
      journal = {\aap},
         year = 2006,
        month = jul,
       volume = {453},
       number = {2},
        pages = {L29-L33},
          doi = {10.1051/0004-6361:20065155},
archivePrefix = {arXiv},
       eprint = {astro-ph/0605353},
 primaryClass = {astro-ph},
       adsurl = {https://ui.adsabs.harvard.edu/abs/2006A&A...453L..29C}
}

@ARTICLE{Croton_2006,
       author = {{Croton}, Darren J. and {Springel}, Volker and {White}, Simon D.~M. and {De Lucia}, G. and {Frenk}, C.~S. and {Gao}, L. and {Jenkins}, A. and {Kauffmann}, G. and {Navarro}, J.~F. and {Yoshida}, N.},
        title = "{The many lives of active galactic nuclei: cooling flows, black holes and the luminosities and colours of galaxies}",
      journal = {\mnras},
         year = 2006,
        month = jan,
       volume = {365},
       number = {1},
        pages = {11-28},
          doi = {10.1111/j.1365-2966.2005.09675.x},
archivePrefix = {arXiv},
       eprint = {astro-ph/0508046},
 primaryClass = {astro-ph},
       adsurl = {https://ui.adsabs.harvard.edu/abs/2006MNRAS.365...11C}
}

@ARTICLE{Debattista_2006,
       author = {{Debattista}, Victor P. and {Mayer}, Lucio and {Carollo}, C. Marcella and {Moore}, Ben and {Wadsley}, James and {Quinn}, Thomas},
        title = "{The Secular Evolution of Disk Structural Parameters}",
      journal = {\apj},
         year = 2006,
        month = jul,
       volume = {645},
       number = {1},
        pages = {209-227},
          doi = {10.1086/504147},
archivePrefix = {arXiv},
       eprint = {astro-ph/0509310},
 primaryClass = {astro-ph},
       adsurl = {https://ui.adsabs.harvard.edu/abs/2006ApJ...645..209D}
}

@ARTICLE{ForsterSchreiber_2006,
   author = {{F{\"o}rster Schreiber}, N.~M. and {Genzel}, R. and {Lehnert}, M.~D. and 
	{Bouch{\'e}}, N. and {Verma}, A. and {Erb}, D.~K. and {Shapley}, A.~E. and 
	{Steidel}, C.~C. and {Davies}, R. and {Lutz}, D. and {Nesvadba}, N. and 
	{Tacconi}, L.~J. and {Eisenhauer}, F. and {Abuter}, R. and {Gilbert}, A. and 
	{Gillessen}, S. and {Sternberg}, A.},
    title = "{SINFONI Integral Field Spectroscopy of z \~{} 2 UV-selected Galaxies: Rotation Curves and Dynamical Evolution}",
  journal = {\apj},
   eprint = {arXiv:astro-ph/0603559},
     year = 2006,
    month = jul,
   volume = 645,
    pages = {1062-1075},
      doi = {10.1086/504403},
   adsurl = {http://adsabs.harvard.edu/abs/2006ApJ...645.1062F}
}

@ARTICLE{Renzini_2006,
       author = {{Renzini}, Alvio},
        title = "{Stellar Population Diagnostics of Elliptical Galaxy Formation}",
      journal = {\araa},
         year = 2006,
        month = sep,
       volume = {44},
       number = {1},
        pages = {141-192},
          doi = {10.1146/annurev.astro.44.051905.092450},
archivePrefix = {arXiv},
       eprint = {astro-ph/0603479},
 primaryClass = {astro-ph},
       adsurl = {https://ui.adsabs.harvard.edu/abs/2006ARA&A..44..141R}
}

@ARTICLE{Daddi_2007a,
       author = {{Daddi}, E. and {Dickinson}, M. and {Morrison}, G. and {Chary}, R. and {Cimatti}, A. and {Elbaz}, D. and {Frayer}, D. and {Renzini}, A. and {Pope}, A. and {Alexander}, D.~M. and {Bauer}, F.~E. and {Giavalisco}, M. and {Huynh}, M. and {Kurk}, J. and {Mignoli}, M.},
        title = "{Multiwavelength Study of Massive Galaxies at z\raisebox{-0.5ex}\textasciitilde2. I. Star Formation and Galaxy Growth}",
      journal = {\apj},
         year = 2007,
        month = nov,
       volume = {670},
       number = {1},
        pages = {156-172},
          doi = {10.1086/521818},
archivePrefix = {arXiv},
       eprint = {0705.2831},
 primaryClass = {astro-ph},
       adsurl = {https://ui.adsabs.harvard.edu/abs/2007ApJ...670..156D}
}

@ARTICLE{Daddi_2007b,
   author = {{Daddi}, E. and {Alexander}, D.~M. and {Dickinson}, M. and {Gilli}, R. and 
	{Renzini}, A. and {Elbaz}, D. and {Cimatti}, A. and {Chary}, R. and 
	{Frayer}, D. and {Bauer}, F.~E. and {Brandt}, W.~N. and {Giavalisco}, M. and 
	{Grogin}, N.~A. and {Huynh}, M. and {Kurk}, J. and {Mignoli}, M. and 
	{Morrison}, G. and {Pope}, A. and {Ravindranath}, S.},
    title = "{Multiwavelength Study of Massive Galaxies at z\~{}2. II. Widespread Compton-thick Active Galactic Nuclei and the Concurrent Growth of Black Holes and Bulges}",
  journal = {\apj},
   eprint = {arXiv:0705.2832},
     year = 2007,
    month = nov,
   volume = 670,
    pages = {173-189},
      doi = {10.1086/521820},
   adsurl = {http://adsabs.harvard.edu/abs/2007ApJ...670..173D}
}

@ARTICLE{Elbaz_2007,
   author = {{Elbaz}, D. and {Daddi}, E. and {Le Borgne}, D. and {Dickinson}, M. and 
	{Alexander}, D.~M. and {Chary}, {R.-R.} and {Starck}, {J.-L.} and 
	{Brandt}, W.~N. and {Kitzbichler}, M. and {MacDonald}, E. and 
	{Nonino}, M. and {Popesso}, P. and {Stern}, D. and {Vanzella}, E.
	},
    title = "{The reversal of the star formation-density relation in the distant universe}",
  journal = {\aap},
   eprint = {arXiv:astro-ph/0703653},
     year = 2007,
    month = jun,
   volume = 468,
    pages = {33-48},
      doi = {10.1051/0004-6361:20077525},
   adsurl = {http://adsabs.harvard.edu/abs/2007A%26A...468...33E}
}

@ARTICLE{Noeske_2007,
   author = {{Noeske}, K.~G. and {Weiner}, B.~J. and {Faber}, S.~M. and {Papovich}, C. and 
	{Koo}, D.~C. and {Somerville}, R.~S. and {Bundy}, K. and {Conselice}, C.~J. and 
	{Newman}, J.~A. and {Schiminovich}, D. and {Le Floc'h}, E. and 
	{Coil}, A.~L. and {Rieke}, G.~H. and {Lotz}, J.~M. and {Primack}, J.~R. and 
	{Barmby}, P. and {Cooper}, M.~C. and {Davis}, M. and {Ellis}, R.~S. and 
	{Fazio}, G.~G. and {Guhathakurta}, P. and {Huang}, J. and {Kassin}, S.~A. and 
	{Martin}, D.~C. and {Phillips}, A.~C. and {Rich}, R.~M. and 
	{Small}, T.~A. and {Willmer}, C.~N.~A. and {Wilson}, G.},
    title = "{Star Formation in AEGIS Field Galaxies since z=1.1: The Dominance of Gradually Declining Star Formation, and the Main Sequence of Star-forming Galaxies}",
  journal = {\apjl},
   eprint = {arXiv:astro-ph/0701924},
     year = 2007,
    month = may,
   volume = 660,
    pages = {L43-L46},
      doi = {10.1086/517926},
   adsurl = {http://adsabs.harvard.edu/abs/2007ApJ...660L..43N}
}

@article{Fiore_2008,
  author       = {Brusa, Marcella and Fiore, Fabrizio and Santini, Paola and Grazian, Andrea and Comastri, Andrea and Zamorani, Gianni and Hasinger, G{\"u}nther and Merloni, Andrea and Civano, Francesca and Fontana, A. and Mainieri, V.},
  title        = {Black Hole Growth and Starburst Activity at $z = 0.6$--$4$ in the Chandra Deep Field South: Host Galaxy Properties of Obscured AGN},
  journal      = {Astronomy \& Astrophysics},
  volume       = {507},
  number       = {3},
  pages        = {1277--1289},
  year         = {2009},
  doi          = {10.1051/0004-6361/200912261}
}

@ARTICLE{Genzel_2008,
       author = {{Genzel}, R. and {Burkert}, A. and {Bouch{\'e}}, N. and {Cresci}, G. and {F{\"o}rster Schreiber}, N.~M. and {Shapley}, A. and {Shapiro}, K. and {Tacconi}, L.~J. and {Buschkamp}, P. and {Cimatti}, A. and {Daddi}, E. and {Davies}, R. and {Eisenhauer}, F. and {Erb}, D.~K. and {Genel}, S. and {Gerhard}, O. and {Hicks}, E. and {Lutz}, D. and {Naab}, T. and {Ott}, T. and {Rabien}, S. and {Renzini}, A. and {Steidel}, C.~C. and {Sternberg}, A. and {Lilly}, S.~J.},
        title = "{From Rings to Bulges: Evidence for Rapid Secular Galaxy Evolution at z \raisebox{-0.5ex}\textasciitilde 2 from Integral Field Spectroscopy in the SINS Survey}",
      journal = {\apj},
         year = 2008,
        month = nov,
       volume = {687},
       number = {1},
        pages = {59-77},
          doi = {10.1086/591840},
archivePrefix = {arXiv},
       eprint = {0807.1184},
 primaryClass = {astro-ph},
       adsurl = {https://ui.adsabs.harvard.edu/abs/2008ApJ...687...59G}
}

@article{vanDokkum_2008,
  author    = {van Dokkum, P.~G. and Franx, M. and Kriek, M. and Holden, B. P. and Labb\'{e}, I. and Moorwood, A. F. M. and Rix, H.-W. and Rudnick, G. and Taylor, E. N. and Toft, S. and van der Wel, A. and Wuyts, S.},
  title     = {Confirmation of the Remarkable Compactness of Massive Quiescent Galaxies at $z \sim 2.3$: Early-Type Galaxies Did Not Form in a Simple Monolithic Collapse},
  journal   = {The Astrophysical Journal Letters},
  volume    = {677},
  pages     = {L5--L8},
  year      = {2008},
  doi       = {10.1086/587957},
  adsurl    = {https://ui.adsabs.harvard.edu/abs/2008ApJ...677L...5V}
}

@article{Brusa_2009,
  author       = {Brusa, Marcella and Fiore, Fabrizio and Santini, Paola and Grazian, Andrea and Comastri, Andrea and Zamorani, Gianni and Hasinger, G{\"u}nther and Merloni, Andrea and Civano, Francesca and Fontana, A. and Mainieri, V.},
  title        = {Black Hole Growth and Starburst Activity at $z = 0.6$--$4$ in the Chandra Deep Field South},
  journal      = {Astronomy \& Astrophysics},
  volume       = {507},
  pages        = {1277--1289},
  year         = {2009},
  doi          = {10.1051/0004-6361/200912261}
}

@ARTICLE{Dekel_2009a,
       author = {{Dekel}, A. and {Birnboim}, Y. and {Engel}, G. and {Freundlich}, J. and {Goerdt}, T. and {Mumcuoglu}, M. and {Neistein}, E. and {Pichon}, C. and {Teyssier}, R. and {Zinger}, E.},
        title = "{Cold streams in early massive hot haloes as the main mode of galaxy formation}",
      journal = {\nat},
         year = 2009,
        month = jan,
       volume = {457},
       number = {7228},
        pages = {451-454},
          doi = {10.1038/nature07648},
archivePrefix = {arXiv},
       eprint = {0808.0553},
 primaryClass = {astro-ph},
       adsurl = {https://ui.adsabs.harvard.edu/abs/2009Natur.457..451D}
}

@article{Dekel_2009b,
doi = {10.1088/0004-637X/703/1/785},
url = {https://doi.org/10.1088/0004-637X/703/1/785},
year = {2009},
month = {sep},
publisher = {The American Astronomical Society},
volume = {703},
number = {1},
pages = {785},
author = {Dekel, Avishai and Sari, Re'em and Ceverino, Daniel},
title = {FORMATION OF MASSIVE GALAXIES AT HIGH REDSHIFT: COLD STREAMS, CLUMPY DISKS, AND COMPACT SPHEROIDS},
journal = {The Astrophysical Journal}
}

@ARTICLE{ForsterSchreiber_2009,
   author = {{F{\"o}rster Schreiber}, N.~M. and {Genzel}, R. and {Bouch{\'e}}, N. and 
	{Cresci}, G. and {Davies}, R. and {Buschkamp}, P. and {Shapiro}, K. and 
	{Tacconi}, L.~J. and {Hicks}, E.~K.~S. and {Genel}, S. and {Shapley}, A.~E. and 
	{Erb}, D.~K. and {Steidel}, C.~C. and {Lutz}, D. and {Eisenhauer}, F. and 
	{Gillessen}, S. and {Sternberg}, A. and {Renzini}, A. and {Cimatti}, A. and 
	{Daddi}, E. and {Kurk}, J. and {Lilly}, S. and {Kong}, X. and 
	{Lehnert}, M.~D. and {Nesvadba}, N. and {Verma}, A. and {McCracken}, H. and 
	{Arimoto}, N. and {Mignoli}, M. and {Onodera}, M.},
    title = "{The SINS Survey: SINFONI Integral Field Spectroscopy of z \~{} 2 Star-forming Galaxies}",
  journal = {\apj},
archivePrefix = "arXiv",
   eprint = {0903.1872},
 primaryClass = "astro-ph.CO",
     year = 2009,
    month = dec,
   volume = 706,
    pages = {1364-1428},
      doi = {10.1088/0004-637X/706/2/1364},
   adsurl = {http://adsabs.harvard.edu/abs/2009ApJ...706.1364F}
}

@ARTICLE{Martig_2009,
   author = {{Martig}, M. and {Bournaud}, F. and {Teyssier}, R. and {Dekel}, A.
	},
    title = "{Morphological Quenching of Star Formation: Making Early-Type Galaxies Red}",
  journal = {\apj},
archivePrefix = "arXiv",
   eprint = {0905.4669},
 primaryClass = "astro-ph.CO",
     year = 2009,
    month = dec,
   volume = 707,
    pages = {250-267},
      doi = {10.1088/0004-637X/707/1/250},
   adsurl = {http://adsabs.harvard.edu/abs/2009ApJ...707..250M}
}

@ARTICLE{Wild_2009,
       author = {{Wild}, Vivienne and {Walcher}, C. Jakob and {Johansson}, Peter H. and {Tresse}, Laurence and {Charlot}, St{\'e}phane and {Pollo}, Agnieszka and {Le F{\`e}vre}, Olivier and {de Ravel}, Loic},
        title = "{Post-starburst galaxies: more than just an interesting curiosity}",
      journal = {\mnras},
         year = 2009,
        month = may,
       volume = {395},
       number = {1},
        pages = {144-159},
          doi = {10.1111/j.1365-2966.2009.14537.x},
archivePrefix = {arXiv},
       eprint = {0810.5122},
 primaryClass = {astro-ph},
       adsurl = {https://ui.adsabs.harvard.edu/abs/2009MNRAS.395..144W}
}

@ARTICLE{PengY_2010,
   author = {{Peng}, {Y.-j.} and {Lilly}, S.~J. and {Kova{\v c}}, K. and 
	{Bolzonella}, M. and {Pozzetti}, L. and {Renzini}, A. and {Zamorani}, G. and 
	{Ilbert}, O. and {Knobel}, C. and {Iovino}, A. and {Maier}, C. and 
	{Cucciati}, O. and {Tasca}, L. and {Carollo}, C.~M. and {Silverman}, J. and 
	{Kampczyk}, P. and {de Ravel}, L. and {Sanders}, D. and {Scoville}, N. and 
	{Contini}, T. and {Mainieri}, V. and {Scodeggio}, M. and {Kneib}, {J.-P.} and 
	{Le F{\`e}vre}, O. and {Bardelli}, S. and {Bongiorno}, A. and 
	{Caputi}, K. and {Coppa}, G. and {de la Torre}, S. and {Franzetti}, P. and 
	{Garilli}, B. and {Lamareille}, F. and {Le Borgne}, {J.-F.} and 
	{Le Brun}, V. and {Mignoli}, M. and {Perez Montero}, E. and 
	{Pello}, R. and {Ricciardelli}, E. and {Tanaka}, M. and {Tresse}, L. and 
	{Vergani}, D. and {Welikala}, N. and {Zucca}, E. and {Oesch}, P. and 
	{Abbas}, U. and {Barnes}, L. and {Bordoloi}, R. and {Bottini}, D. and 
	{Cappi}, A. and {Cassata}, P. and {Cimatti}, A. and {Fumana}, M. and 
	{Hasinger}, G. and {Koekemoer}, A. and {Leauthaud}, A. and {Maccagni}, D. and 
	{Marinoni}, C. and {McCracken}, H. and {Memeo}, P. and {Meneux}, B. and 
	{Nair}, P. and {Porciani}, C. and {Presotto}, V. and {Scaramella}, R.
	},
    title = "{Mass and Environment as Drivers of Galaxy Evolution in SDSS and zCOSMOS and the Origin of the Schechter Function}",
  journal = {\apj},
archivePrefix = "arXiv",
   eprint = {1003.4747},
 primaryClass = "astro-ph.CO",
     year = 2010,
    month = sep,
   volume = 721,
    pages = {193-221},
      doi = {10.1088/0004-637X/721/1/193},
   adsurl = {http://adsabs.harvard.edu/abs/2010ApJ...721..193P}
}

@ARTICLE{Genzel_2011,
       author = {{Genzel}, R. and {Newman}, S. and {Jones}, T. and {F{\"o}rster Schreiber}, N.~M. and {Shapiro}, K. and {Genel}, S. and {Lilly}, S.~J. and {Renzini}, A. and {Tacconi}, L.~J. and {Bouch{\'e}}, N. and {Burkert}, A. and {Cresci}, G. and {Buschkamp}, P. and {Carollo}, C.~M. and {Ceverino}, D. and {Davies}, R. and {Dekel}, A. and {Eisenhauer}, F. and {Hicks}, E. and {Kurk}, J. and {Lutz}, D. and {Mancini}, C. and {Naab}, T. and {Peng}, Y. and {Sternberg}, A. and {Vergani}, D. and {Zamorani}, G.},
        title = "{The Sins Survey of z \raisebox{-0.5ex}\textasciitilde 2 Galaxy Kinematics: Properties of the Giant Star-forming Clumps}",
      journal = {\apj},
         year = 2011,
        month = jun,
       volume = {733},
       number = {2},
          eid = {101},
        pages = {101},
          doi = {10.1088/0004-637X/733/2/101},
archivePrefix = {arXiv},
       eprint = {1011.5360},
 primaryClass = {astro-ph.CO},
       adsurl = {https://ui.adsabs.harvard.edu/abs/2011ApJ...733..101G}
}

@ARTICLE{Saracco_2011,
       author = {{Saracco}, P. and {Longhetti}, M. and {Gargiulo}, A.},
        title = "{Constraining the star formation and the assembly histories of normal and compact early-type galaxies at 1 < z < 2}",
      journal = {\mnras},
         year = 2011,
        month = apr,
       volume = {412},
       number = {4},
        pages = {2707-2716},
          doi = {10.1111/j.1365-2966.2010.18098.x},
archivePrefix = {arXiv},
       eprint = {1011.5740},
 primaryClass = {astro-ph.CO},
       adsurl = {https://ui.adsabs.harvard.edu/abs/2011MNRAS.412.2707S}
}

@article{Simard_2011,
  author    = {Simard, L. and Mendel, J. T. and Patton, D. R. and Ellison, S. L. and McConnachie, A. W.},
  title     = {A Catalog of Bulge+Disk Decompositions and Updated Photometry for 1.12 Million Galaxies in SDSS DR7},
  journal   = {Astrophysical Journal Supplement Series},
  volume    = {196},
  number    = {1},
  pages     = {11},
  year      = {2011},
  doi       = {10.1088/0067-0049/196/1/11},
  adsurl    = {https://ui.adsabs.harvard.edu/abs/2011ApJS..196...11S}
}

@ARTICLE{Wuyts_2011,
   author = {{Wuyts}, S. and {F{\"o}rster Schreiber}, N.~M. and {van der Wel}, A. and 
	{Magnelli}, B. and {Guo}, Y. and {Genzel}, R. and {Lutz}, D. and 
	{Aussel}, H. and {Barro}, G. and {Berta}, S. and {Cava}, A. and 
	{Graci{\'a}-Carpio}, J. and {Hathi}, N.~P. and {Huang}, K.-H. and 
	{Kocevski}, D.~D. and {Koekemoer}, A.~M. and {Lee}, K.-S. and 
	{Le Floc'h}, E. and {McGrath}, E.~J. and {Nordon}, R. and {Popesso}, P. and 
	{Pozzi}, F. and {Riguccini}, L. and {Rodighiero}, G. and {Saintonge}, A. and 
	{Tacconi}, L.},
    title = "{Galaxy Structure and Mode of Star Formation in the SFR-Mass Plane from z \~{} 2.5 to z \~{} 0.1}",
  journal = {\apj},
archivePrefix = "arXiv",
   eprint = {1107.0317},
 primaryClass = "astro-ph.CO",
     year = 2011,
    month = dec,
   volume = 742,
      eid = {96},
    pages = {96},
      doi = {10.1088/0004-637X/742/2/96},
   adsurl = {http://adsabs.harvard.edu/abs/2011ApJ...742...96W}
}

@ARTICLE{Bruce_2012,
       author = {{Bruce}, V.~A. and {Dunlop}, J.~S. and {Cirasuolo}, M. and {McLure}, R.~J. and {Targett}, T.~A. and {Bell}, E.~F. and {Croton}, D.~J. and {Dekel}, A. and {Faber}, S.~M. and {Ferguson}, H.~C. and {Grogin}, N.~A. and {Kocevski}, D.~D. and {Koekemoer}, A.~M. and {Koo}, D.~C. and {Lai}, K. and {Lotz}, J.~M. and {McGrath}, E.~J. and {Newman}, J.~A. and {van der Wel}, A.},
        title = "{The morphologies of massive galaxies at 1 < z < 3 in the CANDELS-UDS field: compact bulges, and the rise and fall of massive discs}",
      journal = {\mnras},
         year = 2012,
        month = dec,
       volume = {427},
       number = {2},
        pages = {1666-1701},
          doi = {10.1111/j.1365-2966.2012.22087.x},
archivePrefix = {arXiv},
       eprint = {1206.4322},
 primaryClass = {astro-ph.CO},
       adsurl = {https://ui.adsabs.harvard.edu/abs/2012MNRAS.427.1666B}
}

@ARTICLE{Peng_2012,
       author = {{Peng}, Ying-jie and {Lilly}, Simon J. and {Renzini}, Alvio and {Carollo}, Marcella},
        title = "{Mass and Environment as Drivers of Galaxy Evolution. II. The Quenching of Satellite Galaxies as the Origin of Environmental Effects}",
      journal = {\apj},
         year = 2012,
        month = sep,
       volume = {757},
       number = {1},
          eid = {4},
        pages = {4},
          doi = {10.1088/0004-637X/757/1/4},
archivePrefix = {arXiv},
       eprint = {1106.2546},
 primaryClass = {astro-ph.CO},
       adsurl = {https://ui.adsabs.harvard.edu/abs/2012ApJ...757....4P}
}

@article{Ilbert_2013,
  author    = {Ilbert, O. and McCracken, H. J. and Le F\`{e}vre, O. and Capak, P. and Dunlop, J. and Karim, A. and Renzini, A. and Caputi, K. and Boissier, S. and Arnouts, S. and others},
  title     = {Mass assembly in quiescent and star-forming galaxies since $z \sim 4$: results from the UltraVISTA Survey},
  journal   = {Astronomy \& Astrophysics},
  volume    = {556},
  pages     = {A55},
  year      = {2013},
  doi       = {10.1051/0004-6361/201321621},
  adsurl    = {https://ui.adsabs.harvard.edu/abs/2013A&A...556A..55I}
}

@ARTICLE{Kormendy_Ho_2013,
       author = {{Kormendy}, John and {Ho}, Luis C.},
        title = "{Coevolution (Or Not) of Supermassive Black Holes and Host Galaxies}",
      journal = {\araa},
         year = 2013,
        month = aug,
       volume = {51},
       number = {1},
        pages = {511-653},
          doi = {10.1146/annurev-astro-082708-101811},
archivePrefix = {arXiv},
       eprint = {1304.7762},
 primaryClass = {astro-ph.CO},
       adsurl = {https://ui.adsabs.harvard.edu/abs/2013ARA&A..51..511K}
}

@ARTICLE{Bluck_2014,
       author = {{Bluck}, Asa F.~L. and {Mendel}, J. Trevor and {Ellison}, Sara L. and {Moreno}, Jorge and {Simard}, Luc and {Patton}, David R. and {Starkenburg}, Else},
        title = "{Bulge mass is king: the dominant role of the bulge in determining the fraction of passive galaxies in the Sloan Digital Sky Survey}",
      journal = {\mnras},
         year = 2014,
        month = jun,
       volume = {441},
       number = {1},
        pages = {599-629},
          doi = {10.1093/mnras/stu594},
archivePrefix = {arXiv},
       eprint = {1403.5269},
 primaryClass = {astro-ph.GA},
       adsurl = {https://ui.adsabs.harvard.edu/abs/2014MNRAS.441..599B}
}

@ARTICLE{Dekel_2014,
       author = {{Dekel}, A. and {Burkert}, A.},
        title = "{Wet disc contraction to galactic blue nuggets and quenching to red nuggets}",
      journal = {\mnras},
         year = 2014,
        month = feb,
       volume = {438},
       number = {2},
        pages = {1870-1879},
          doi = {10.1093/mnras/stt2331},
archivePrefix = {arXiv},
       eprint = {1310.1074},
 primaryClass = {astro-ph.CO},
       adsurl = {https://ui.adsabs.harvard.edu/abs/2014MNRAS.438.1870D}
}

@ARTICLE{ForsterSchreiber_2014,
       author = {{F{\"o}rster Schreiber}, N.~M. and {Genzel}, R. and {Newman}, S.~F. and {Kurk}, J.~D. and {Lutz}, D. and {Tacconi}, L.~J. and {Wuyts}, S. and {Bandara}, K. and {Burkert}, A. and {Buschkamp}, P. and {Carollo}, C.~M. and {Cresci}, G. and {Daddi}, E. and {Davies}, R. and {Eisenhauer}, F. and {Hicks}, E.~K.~S. and {Lang}, P. and {Lilly}, S.~J. and {Mainieri}, V. and {Mancini}, C. and {Naab}, T. and {Peng}, Y. and {Renzini}, A. and {Rosario}, D. and {Shapiro Griffin}, K. and {Shapley}, A.~E. and {Sternberg}, A. and {Tacchella}, S. and {Vergani}, D. and {Wisnioski}, E. and {Wuyts}, E. and {Zamorani}, G.},
        title = "{The Sins/zC-Sinf Survey of z \raisebox{-0.5ex}\textasciitilde 2 Galaxy Kinematics: Evidence for Powerful Active Galactic Nucleus-Driven Nuclear Outflows in Massive Star-Forming Galaxies}",
      journal = {\apj},
         year = 2014,
        month = may,
       volume = {787},
       number = {1},
          eid = {38},
        pages = {38},
          doi = {10.1088/0004-637X/787/1/38},
archivePrefix = {arXiv},
       eprint = {1311.2596},
 primaryClass = {astro-ph.CO},
       adsurl = {https://ui.adsabs.harvard.edu/abs/2014ApJ...787...38F}
}

@ARTICLE{Lang_2014,
       author = {{Lang}, Philipp and {Wuyts}, Stijn and {Somerville}, Rachel S. and {F{\"o}rster Schreiber}, Natascha M. and {Genzel}, Reinhard and {Bell}, Eric F. and {Brammer}, Gabe and {Dekel}, Avishai and {Faber}, Sandra M. and {Ferguson}, Henry C. and {Grogin}, Norman A. and {Kocevski}, Dale D. and {Koekemoer}, Anton M. and {Lutz}, Dieter and {McGrath}, Elizabeth J. and {Momcheva}, Ivelina and {Nelson}, Erica J. and {Primack}, Joel R. and {Rosario}, David J. and {Skelton}, Rosalind E. and {Tacconi}, Linda J. and {van Dokkum}, Pieter G. and {Whitaker}, Katherine E.},
        title = "{Bulge Growth and Quenching since z = 2.5 in CANDELS/3D-HST}",
      journal = {\apj},
         year = 2014,
        month = jun,
       volume = {788},
       number = {1},
          eid = {11},
        pages = {11},
          doi = {10.1088/0004-637X/788/1/11},
archivePrefix = {arXiv},
       eprint = {1402.0866},
 primaryClass = {astro-ph.GA},
       adsurl = {https://ui.adsabs.harvard.edu/abs/2014ApJ...788...11L}
}

@ARTICLE{Madau_2014,
       author = {{Madau}, Piero and {Dickinson}, Mark},
        title = "{Cosmic Star-Formation History}",
      journal = {\araa},
         year = 2014,
        month = aug,
       volume = {52},
        pages = {415-486},
          doi = {10.1146/annurev-astro-081811-125615},
archivePrefix = {arXiv},
       eprint = {1403.0007},
 primaryClass = {astro-ph.CO},
       adsurl = {https://ui.adsabs.harvard.edu/abs/2014ARA&A..52..415M}
}

@ARTICLE{Omand_2014,
       author = {{Omand}, Conor M.~B. and {Balogh}, Michael L. and {Poggianti}, Bianca M.},
        title = "{The connection between galaxy structure and quenching efficiency}",
      journal = {\mnras},
         year = 2014,
        month = may,
       volume = {440},
       number = {1},
        pages = {843-858},
          doi = {10.1093/mnras/stu331},
archivePrefix = {arXiv},
       eprint = {1402.3394},
 primaryClass = {astro-ph.GA},
       adsurl = {https://ui.adsabs.harvard.edu/abs/2014MNRAS.440..843O}
}

@ARTICLE{Salim_2014,
       author = {{Salim}, S.},
        title = "{Green Valley Galaxies}",
      journal = {Serbian Astronomical Journal},
         year = 2014,
        month = dec,
       volume = {189},
        pages = {1-14},
          doi = {10.2298/SAJ1489001S},
archivePrefix = {arXiv},
       eprint = {1501.01963},
 primaryClass = {astro-ph.GA},
       adsurl = {https://ui.adsabs.harvard.edu/abs/2014SerAJ.189....1S}
}

@ARTICLE{Speagle_2014,
   author = {{Speagle}, J.~S. and {Steinhardt}, C.~L. and {Capak}, P.~L. and 
	{Silverman}, J.~D.},
    title = "{A Highly Consistent Framework for the Evolution of the Star-Forming ``Main Sequence'' from z \~{} 0-6}",
  journal = {\apjs},
archivePrefix = "arXiv",
   eprint = {1405.2041},
     year = 2014,
    month = oct,
   volume = 214,
      eid = {15},
    pages = {15},
      doi = {10.1088/0067-0049/214/2/15},
   adsurl = {http://adsabs.harvard.edu/abs/2014ApJS..214...15S}
}

@article{Straatman_2014,
  author    = {Straatman, C.~M.~S. and Labb\'{e}, I. and Spitler, L.~R. and Glazebrook, K. and Quadri, R.~F. and Brammer, G.~B. and Whitaker, K.~E. and Franx, M. and Kriek, M. and Marchesini, D. and Rudnick, G. and van Dokkum, P.~G.},
  title     = {The NMBS\footnote{NOAO (Northern) Mixed Band Survey} Large Quiescent Galaxy Sample: Number Density, Mass Function, and Stellar Age Distribution at $0.4 < z < 2.2$},
  journal   = {Astrophysical Journal},
  volume    = {783},
  number    = {2},
  pages     = {L14},
  year      = {2014},
  doi       = {10.1088/2041-8205/783/2/L14},
  adsurl    = {https://ui.adsabs.harvard.edu/abs/2014ApJ...783L..14S}
}


\end{document}